\def\mnras{Mon. Not. R. Astr. Soc.}  % Monthly Notices of the RAS
\def\apj{Astrophys. J.}  
\def\araa{Ann. Rev. of A. \& A.}  
\input{aipcheck}

\documentclass[
    ,final            % use final for the camera ready runs
  ]
  {aipproc}

\layoutstyle{6x9}

\begin{document}

\title{Cosmological pseudobulge formation}


\author{Takashi Okamoto}{
  address={Center for Computational Sciences, University of Tsukuba, 1-1-1 Tennodai, Tsukuba 305-8577 Ibaraki, Japan}
}

\begin{abstract}
Bulges can be classified into classical and pseudobulges; the former
are considered to be end products of galactic mergers and the latter
to form via secular evolution of galactic disks. Observationally,
bulges of disk galaxies are mostly pseudobulges, including the Milky
Way's. We here show, by using self-consistent cosmological
simulations of galaxy formation, that the formation of
pseudobulges of Milky Way-sized disk galaxies has mostly completed 
before disk formation; 
thus the main channel of pseudobulge formation is not secular evolution
of disks. Our pseudobulges form by rapid gas supply at high-redshift
and their progenitors would be observed as high-redshift disks.
\end{abstract}

\maketitle

%%%%%%%%%%%%%%%%%%%%%%%%%%%%%%%%%%%%%%%%%%%%
%% MAINMATTER
%%%%%%%%%%%%%%%%%%%%%%%%%%%%%%%%%%%%%%%%%%%%

\section{Introduction}

Bulges of disk galaxies can be classified into two types: 
classical bulges and pseudobulges. 
Pseudobulges show substantial rotational support and disky 
or boxy/peanut edge-on isophoto shapes. 
Pseudobulges are thus thought to have formed via secular 
evolution of discs \cite{kk04}. 
The standard picture of galaxy formation predicts that 
galaxies from through hierarchical merging that naturally 
produces classical bulges. However, more than half of 
bulges of nearby large disk galaxies are pseudobulges
\citep{weinzirl09, kormendy10}. 
Therefore the high frequency of pseudobulges could be 
a challenge to the standard cosmology. 

Investigating formation processes of simulated bulges of 
disk galaxies should provide important clues for understanding 
of bulge formation. Such simulations must resolve detailed 
structure of galaxies such as shape of bulges. 
Only recently cosmological simulations with high enough 
resolution have become possible \citep{of09, ofjt10, eris}.

In this paper, we analyze the bulges of two Milky Way-sized 
galaxies formed in $\Lambda$CDM simulations; the initial 
conditions are selected from the Aquarius project
\citep{aquarius}: `Aq-C' and `Aq-D' in their labeling system.  
The simulation presented in this paper include a number of 
baryonic processes known to be relevant to galaxy formation. 
We make use of a model that has already had success in 
reproducing properties of the Local Group satellite galaxies
\citep{of09, ofjt10}. The details of the simulations please 
refer to \citet{okamoto12}.

\section{Results}

\begin{figure}
  \includegraphics[height=.2\textheight]{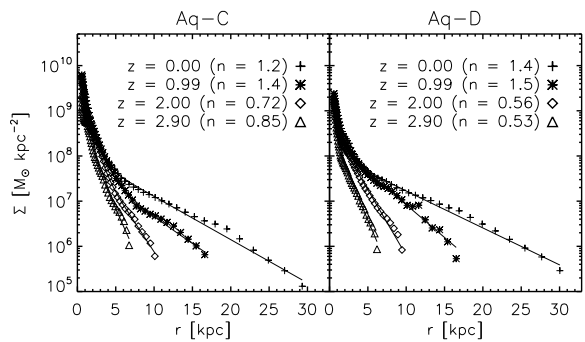}
  \caption{Evolution of the surface stellar density profiles. 
  The surface stellar density profiles of Aq-C (left) and 
  Aq-D (right) galaxies and their main progenitors are 
  presented. The profiles around redshift 0, 1, 2, and 3 
  are indicated by the plus signs, asterisks, diamonds, and 
  triangles, respectively. 
  Each profile is fitted by a combination of a S\'ersic bulge 
  and an exponential disk, which is shown by a solid line. 
  The S\'ersic indices are shown in each panel. 
}
\end{figure}
In order to classify bulges, the S\'ersic profile fitting 
for a surface stellar density profile is frequently used: 
\begin{equation}
  \Sigma(r) = \Sigma_{\rm e} \exp \left[
    -b_n 
    \left\{
      \left(\frac{r}{R_{\rm e}}\right)^\frac{1}{n} - 1.0
    \right\}
  \right], 
\end{equation}
where $R_{\rm e}$ is the effective radius and $\Sigma_{\rm e}$ is 
the surface density at this radius, respectively, and $n$ is the 
S\'ersic index. The parameter $b_{n}$ is well approximated by 
$b_n = 2 n - 0.324$. 
Bulges with $n < 2$ are usually classified as pseudobulges. 
We show the evolution of the surface stellar density profiles 
in Fig.~1. We fit each profile by a combination of the S\'ersic 
and exponential profiles. 

We obtain $n \simeq 1.2$ and $1.4$ for Aq-C and -D's bulges, 
respectively; thus both bulges have pseudobulge-like profiles. 
We have confirmed that they not only show pseudobulge-like 
profiles, but they also have pseudobulge-like `disky' bulge 
shape \citep[see][]{okamoto12}. 

\begin{figure}
  \includegraphics[height=.2\textheight]{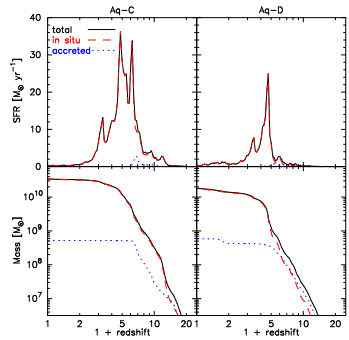}
  \caption{Distribution of formation times of bulge stars 
    expressed in terms of redshift. Stars lie within 3~kpc 
    from the galaxy center at redshift 0 are identified as 
    the bulge stars. The solid lines indicate all the bulge 
    stars, while the dashed and dotted lines respectively 
    indicate those born in the main progenitors (in situ) 
    and those brought by satellites (accreted). 
    The upper and lower panels show the same data in differential 
    and cumulative form, respectively. 
}
\end{figure}

In Fig.~2, we show the formation histories of bulges stars, 
where we distinguish between stars born in the main progenitors 
(in situ) and those brought by accreted satellites (accreted). 
As shown in the top panels, most of bulge stars are born in 
starbursts between redshift 2 and 6 in Aq-C and between 2 and 4 
in Aq-D. 
Thus, mergers do not contribute to the formation of the bulges. 
There is little star formation activity below redshift 1 in the 
bulge of Aq-C; Aq-D's bulge show slight activity at this epoch. 
About half of the star formation in the bulge of Aq-D below 
redshift 1 in the gas clumps. The contribution from the clumps is
only about 10\%. 

\subsection{Conclusion}

We have performed $N$-body/SPH cosmological simulations of 
galaxy formation, in which two Milky Way-sized galaxy have 
formed: Aq-C and Aq-D. Both galaxies have well-defined disks 
with the bulge-to-disk mass ratio, $B/T \simeq 0.6$ and 0.3 at 
redshift 0. 
The S\'sersic indices for Aq-C's and Aq-D's bulges are 
1.2 and 1.4, respectively. These values suggest that both 
bulges are pseudobulges. 
The pseudobulges mainly form by high-redshift starbursts 
before redshift 2. 
The evolution of the surface stellar density profiles reveals 
that the pseudobulge components are already in place at redshift 
2--3 as disky components with small scale length. 
The mass of these components at redshift 2 accounts for $\sim 70$\% and 
$\sim 87$\% of the final pseudobulge mass of Aq-C and Aq-D, 
respectively. 
These progenitors of the pseudobulges would be observed as high-redshift 
disks. The formation scenario of pseudobulges by high-redshift starbursts 
provides an explanation of pseudobulges in early-type disk galaxies 
such as S0 and Sa galaxies. 
Pseudobulges do exist in early-type disks and secular evolution may take 
too long to form such large pseudobulges.

%%%%%%%%%%%%%%%%%%%%%%%%%%%%%%%%%%%%%%%%%%%%%%%%
%% BACKMATTER
%%%%%%%%%%%%%%%%%%%%%%%%%%%%%%%%%%%%%%%%%%%%%%%%

\begin{theacknowledgments}
  The simulations were performed with T2K-Tsukuba at Center 
  for Computational Sciences in University of Tsukuba. 
  This work was supported by Grant-in-Aid for Young Scientists 
  (B: 24740112) and by MEXT HPCI STRATEGIC PROGRAM. 
\end{theacknowledgments}



\bibliographystyle{aipproc}   % if natbib is available

%\bibliography{okamoto_ref}

%  \typeout{******************************************}
%  \typeout{** Please run "bibtex \jobname" to optain}
%  \typeout{** the bibliography and then re-run LaTeX}
%  \typeout{** twice to fix the references!}
%  \typeout{******************************************}

\end{document}